\documentstyle[eqsecnum,amssymb,multicol,pre,aps]{revtex}

\newcommand{\rvec}{{\bf r}}
		
\newcommand{\la}{\left<}		
\newcommand{\ra}{\right>}		

\input epsf

\begin{document}
\enlargethispage*{\baselineskip}
\title{BROWNIAN MOTION AND POLYMER STATISTICS
ON CERTAIN CURVED MANIFOLDS}
\author{Radu P. Mondescu\cite{byline} and M. Muthukumar} \address{Department of
Physics \& Astronomy, and Polymer Science \& Engineering Department and
Materials Research Science and Engineering Center, \\ University of
Massachusetts, Amherst, MA 01003}
\maketitle
\tighten
\begin{abstract}
In this paper we have considered a Gaussian polymer chain of length $L$ as an
intrinsic object enclosed on a surface embedded in the Euclidean space. When
the surfaces are the sphere $\text{S}^{D-1}$ in $D$ dimensions and the
cylinder, the cone and the curved torus in ${\Bbb R}^3$, we have calculated
analytically and numerically (using the diffusion equation and the
path-integral approach) the probability distribution function $\text{G}({\bf
R}|{\bf R}'; L)$ of the end-to-end vector ${\bf R} - {\bf R}'$ and the
mean-square end-to-end distance $\la ({\bf R}-{\bf R}')^2 \ra$ of the polymer
chain. Our findings are: the curvature of the surfaces induces a {\sf
geometrical localization area}; at short scales ($L\to 0$) the polymer is {\sf
locally flat} and the mean-square end-to-end distance is just the Gaussian
value $L l$ ($l$ = Kuhn length), independent of the metric properties of the
surface; at large scales ($L\to\infty$), ${\la ({\bf R}-{\bf R}')^2 \ra}$ tends
to a constant value in the sphere case, it is linear in $L$ for the cylinder
and reaches different constant values (as a function of the geometry of the
surface) for the curved torus. In the case of the cone, contraction of the
chain is induced at all length scales by the presence of the vertex, as a
function of the opening angle $2 \alpha$ and the end position ${\bf R}'$ of the
chain. {Explicit crossover formulas are derived for $\text{G}({\bf
R}|{\bf R}'; L)$ and $\la ({\bf R}-{\bf R}')^2 \ra$.}
\end{abstract}
\pacs{36.20.Ey,05.20.-y,05.40.+j}

\begin{multicols}{2}
\section{INTRODUCTION} 

Statistics and topological characteristics of polymers embedded in
curved interfaces are of significance in biological problems related to
the way polymers wrap around spherical and rodlike macromolecules and
vesicles, and the role played by geometry and topology in the dynamics
of a biomolecule captured on the membrane of a cell \cite{NEW}. These
types of issues arise in other area of physics as well: the quantum
mechanics of free particles in curved and multiply connected spaces
(e.g. see Ref.~\cite{SCHUL}), the physics of vortex lines in
Type II superconductors \cite{SUPERCOND} and the rotational Brownian
motion.

In realistic situations, excluded-volume interaction between polymer segments
and fluctuations of curved interfaces need to be accounted for. Under these
circumstances, only approximate analytical results can be generally obtained
via perturbation analysis, variational techniques or renormalization group
theory. Before embarking on such an effort, it is necessary to solve the
problem of a free polymer chain, characterized only by its connectivity, on a
curved, fixed surface. 

In the present paper we derive, using path-integral and
spectral representation, the probability distribution of the end-to-end vector
${\bf G}({\bf R} | {\bf R}'; L)$ and the size---expressed by the mean-square
end-to-end distance $\la ({\bf R} - {\bf R}')^2 \ra$ in the ambient space---of
an ideal (non-interacting) Gaussian polymer lying on a curved manifold.

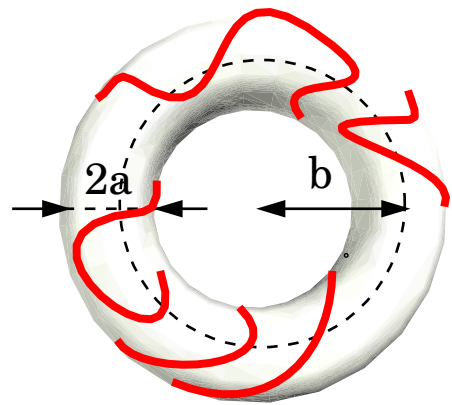
\begin{figure}
\caption{The geometrical data that characterize the sphere (A), the cylinder
(B), the cone (C) and the torus (D): $a$ is the radius of the sphere, of the
cylinder and of the circular cross-section of the torus; $b$ is the radius of
the axial circle of the torus; $\alpha$ is the angle of the cone at the
vertex.}
\label{figs:geometry}
\end{figure}

The curved surfaces studied in this paper are the $\text{S}^{D-1}$ {\sf
sphere} in $D$ dimensions (Fig.~\ref{figs:geometry}A, for $D = 3$); the
{\sf cylinder} (Fig.~\ref{figs:geometry}B), the circular {\sf cone}
(Fig.~\ref{figs:geometry}C) and the {\sf torus}
(Fig.~\ref{figs:geometry}D) in ${\Bbb R}^3$.  The geometrical parameters
are shown in Fig.~\ref{figs:geometry}.  The polymer is an intrinsic
$D\!-\!1$--dimensional object in the spherical surface, while it is
2--dimensional in the other surfaces.

Although, due to the equivalence between the random-walk problem and the
statistics of an ideal Gaussian chain \cite{EDOI}, many results are
already available from the theory of Brownian motion, the specific
problems we address here have not been answered yet.

	In our calculations we have used the path-integral
representation \cite{EDOI,FREED} of the probability distribution
function $\text{G}({\bf R} | {\bf R}';L)$ of a Gaussian polymer chain
and the diffusion equation obeyed by this probability. As mentioned, no
excluded volume or other types of interactions are accounted for in this
paper.
%

\section{FORMULATION}

	Let us consider a $D$--dimensional Gaussian polymer chain with
$N$ links of Kuhn-length $l$ and total length $L = N l$, with the first
bead located at ${\bf R}'$ and the last one at ${\bf R}$.  The
end-to-end vector ${\bf R}-{\bf R}'$ is distributed according to the
following probability distribution function, given as an explicit path
integral\cite{FREED}
\begin{eqnarray}
\text{G}({\bf R} | {\bf R}'; L) & = & \lim_{\stackrel{\scriptstyle
N\to\infty}{\stackrel{\scriptstyle \delta s\to 0}{N\delta s = L}}}
\left({D\over 2\pi l \delta s}\right)^{D N\over 2}\int_{\rvec_0 = {\bf
R}'}^{\rvec_N = {\bf R}} \prod_{j = 1}^{N-1} d^D\rvec_j \, \nonumber\\ &\times
& \exp \left[- {D\over 2 l \delta s}\sum_{j=1}^N (\rvec_j - \rvec_{j-1})^2
\right] ,
\label{path_int}
\end{eqnarray}
where $D$ is the intrinsic dimension of the polymer, $\delta s$ is the
change in the arc-length $s$ along the chain and $\rvec_j$ is the
position vector of the $j$-th bead of the polymer chain.  Note that the
distance on the surface is calculated as $(\rvec_j - \rvec_{j-1})^2 =
g_{uv}(u_j - u_{j-1}) \, (v_j - v_{j-1})$, where $u, v$ are the
intrinsic coordinates and the metric tensor $g_{uv}$ is discretized in a
symmetric form or using the midpoint rule (e.g. see Ref.\
[\onlinecite{KLEINERT,KHANDEKAR}]). In ${\Bbb R}^D$, the expression
(\ref{path_int}) gives the Gaussian probability distribution:
\begin{equation}
\text{G}({\bf R}|{\bf R}'; L) = \left({D\over 2\pi L l}\right)^{D\over 2}
\exp \left[- {D\over 2 L l} ({\bf R} - {\bf R}')^2 \right], 
\end{equation}
The probability $\text{G}({\bf R} | {\bf R}'; L)$ obeys the diffusion
equation
\begin{equation} 
\left({\partial\over \partial L} - {l\over 2 D}\Delta \right)
\text{G}({\bf R} | {\bf R}'; L) = 0 ,
\label{diff_eq}
\end{equation}
subject to the initial condition 
\begin{equation}
\lim_{L\to 0} \text{G}({\bf R} | {\bf R}'; L) = \delta ^{(D)}({\bf R} - {\bf
R}') . 
\label{initial}
\end{equation}
When the polymer lies in a curved manifold, one needs to employ the
Laplace-Beltrami operator given by (e.g. see Ref.\ [\onlinecite{KLEINERT}])
\begin{equation}
\Delta = {1\over \sqrt{g}} \partial_i g^{ij} \partial_j , 
\label{LapBel}
\end{equation}
where $g^{ij}$ is the inverse of the metric tensor $g_{ij}$ of the surface and
$g = \det (g_{ij})$.  We remark that $g_{ij}$ is the induced
metric from the ambient embedding space (${\Bbb R}^D$ or ${\Bbb R}^3$). 

	One can obtain the solution of Eq.~(\ref{diff_eq}) satisfying
the initial condition (\ref{initial}) by applying a Laplace transform
with respect to $L$ that gives the Green's function equation
\begin{equation}
\left( E - {l\over 2 D} \Delta \right)
\widetilde{\text{G}}({\bf R}| {\bf R}'; E) = \delta^{(D)}({\bf R} - {\bf R}'),
\label{Green}
\end{equation}
which can be solved by an eigenfunction expansion of $\widetilde{\text{G}}$
followed by an inverse Laplace transform that eventually yields the {\sf heat
kernel}:
\begin{equation}
\text{G}({\bf R} | {\bf R}'; L) = \sum_{\mu} \phi_{\mu}({\bf R})
\phi_{\mu}^{\ast} ({\bf R}') \exp(- E_{\mu} L).
\label{heat_kernel}
\end{equation}
Here $\phi_{\mu}$ and $E_{\mu}$ are obtained from the eigenvalue equation 
\begin{equation}
- {l\over 2 D}\Delta\phi_{\mu} ({\bf R}) =
E_{\mu} \phi_{\mu} ({\bf R}) ,
\label{eigen} 
\end{equation}
with $\mu$ some generalized index.  In general, the probability distribution
$\text{G}({\bf R} | {\bf R}'; L)$ is not translational invariant. The {\sf
mean-square end-to-end distance} can be calculated from 
\begin{equation}
\la({\mathbf R}- {\mathbf R}')^2\ra = {\int\! d^{D}{\bf R}\,d^{D}{\bf R}' \,
({\mathbf R}-{\mathbf R}')^2 \, \text{G}({\bf R} | {\bf R}'; L)\over \int\!
d^D {\mathbf R} \, d^{D}{\bf R}' \, \text{G}({\bf R} | {\bf R}'; L)} .
\label{end_to_end}
\end{equation}
It is to be noted that the calculated mean-square distance represents the
distance between the ends of the polymer measured by an outside
observer, located in the ambient space, which is different from the
geodesic separation measured along the surface.

\section{CALCULATIONS AND RESULTS} 

Due to the known equivalence \cite{EDOI,WIEGEL} between the diffusion equation 
(\ref{diff_eq}) and the Schr\"odinger equation for a free particle, we can
readily obtain the polymer probability distribution $\text{G}({\bf R}, {\bf
R}'; L)$ for the sphere and the cylinder cases, from the corresponding
probability amplitudes for a rigid rotor in $D\/$ dimensions \cite{GROS} and
for a $2$--$D\/$ free particle moving on the surface of a cylinder,
respectively.

\subsection{\underline{The Sphere $\text{S}^{D-1}$ in ${\Bbb R}^D$}}

Let $a$ be the radius of the $D-1$ dimensional sphere centered at the origin of
the coordinate system. Then ${\bf R}(s)$ is a vector of magnitude $a$ that
measures the position of the bead at $s$ about the center of the sphere. The
intrinsic dimension of the polymer is $D-1$. In spherical coordinates the
Laplace-Beltrami operator (\ref{LapBel}) becomes \cite[\S 4.2.3]{KHANDEKAR}
\begin{equation}
\Delta = {1\over a^2} {\cal L}^2,
\end{equation}
where ${\cal L}^2$ is the $D$-dimensional Legendre operator, satisfying the
eigenvalue equation
\begin{equation}
{\cal L}^2 \text{S}^{m}_J (\Omega) = - J (J + D - 2) \text{S}^{m}_J
 (\Omega).
\label{hyper_eq}
\end{equation}
The $\text{S}_J^{m}$ functions are the hyperspherical harmonics
orthonormal and complete on the unit $\text{S}^{D-1}$ sphere \cite{MO}:
\begin{eqnarray}
\int\!  d\Omega \,\text{S}_J^{m}(\Omega) \text{S}_{J'}^{m '}(\Omega) & =
& \delta_{JJ'}\delta_{m m'} \\ \sum_{J = 0}^{\infty}\!\sum_{m = 1}^M
\text{S}_J^{m}(\Omega) \text{S}_J^{m \ast}(\Omega ') & = & \delta^{(D)}
(\Omega - \Omega ')\nonumber ,
\end{eqnarray}
where $ M = {(2 J + D - 2)\, (J + D - 3)! \over J! (D - 2)!}$ and
$\Omega$ is the solid angle in $D$ dimensions.  Using
Eq.~(\ref{hyper_eq}) and the eigenvalue expansion (\ref{heat_kernel}),
where $\phi_{\mu}\to \text{S}_J^{m}$ and $E_{\mu}\to J (J + D - 2)$, we
readily find the probability distribution function of a polymer chain
(or equivalently, of a random walk) that starts at ${\bf R}'$ and ends
at ${\bf R}$ in $N = L/l$ steps, on the $\text{S}^{D-1}$ sphere:
\begin{eqnarray}
\text{G}({\bf R} | {\bf R}'; L) & = & {1\over a^{D-1}}
\sum_{J=0}^{\infty}\sum_{m = 1 }^{M} \text{S}_J^{m}(\Omega) \,
\text{S}_J^{m\ast}(\Omega ') \label{heat_kernel_sphere}\\ 
& \times & \exp\left[ - {L l \over 2 a^2}{J(J + D -2)\over D-1}\right]
\nonumber .
\end{eqnarray}
The probability found above is normalized properly to one ($\int\! d^D
{\bf R} \, \text{G}({\bf R} | {\bf R}'; L)$ = 1).  In ${\Bbb R}^3$ we
recover the known \cite{ITZYK} rotational diffusion result:
\begin{eqnarray}
\text{G}({\bf R} | {\bf R}'; L) &=&  {1\over a^2}
\sum_{l=0}^{\infty}\sum_{m = -l }^{l} \text{Y}_{lm}(\Omega),
\text{Y}_{lm}^{\ast}(\Omega ') \nonumber\\
& \times& \exp\left[ - {L l \over 4 a^2} J(J + 1) \right].
\label{heat_kernel_sphereR3}
\end{eqnarray}
Because $\text{G}({\bf R} | {\bf R}'; L)$ is translationally invariant,
the end-to-end distance can be calculated as a conditional expectation
value by dropping the integration over ${\bf R}'$ in
Eq.~(\ref{end_to_end}).  To evaluate the remaining integral, we use the
addition theorem for hyper-spherical harmonics \cite{MO}
\begin{eqnarray}
\sum_{m = 1 }^{M} \text{S}_J^{m}(\Omega) \,
\text{S}_J^{m\ast}(\Omega ') &=& {1\over S_D} {(2 J + D - 2)\over D -
2}\nonumber\\
&\times& C_J^{(D - 2)/2}(\cos\Psi_{{\bf R}{\bf R}'}) , 
\end{eqnarray}
where $S_D = {2 \pi^{D/2}\over \Gamma (D/2)}$ and $C_J^{\nu}$ are the
Gegenbauer polynomials of argument $\cos\psi_{{\bf R}{\bf R}'}$.
Writing ${({\bf R} - {\bf R}')^2} = {2 a^2 [1 - \cos\Psi_{{\bf R}{\bf
R}'}]}$ and inserting (\ref{heat_kernel_sphere}) in (\ref{end_to_end}),
we get:
\begin{eqnarray}
\la({\mathbf R}- {\mathbf R}')^2\ra  &=& 2 a^2 \left(1 - {1\over S_D}
\sum_{J=0}^{\infty}\!\int\! d\Omega\, {2 J + D -2\over D - 2}
\right.\nonumber\\
&\times& \cos\Psi_{{\bf R}{\bf R}'} C_J^{(D - 2)/2}(\cos\Psi_{{\bf R}{\bf
R}'})\label{sphere_end}\\
&\times& \left. \exp\left[ - {L l\over 2 a^2} {J (J + D - 2)\over D -
1}\right]\right) .\nonumber
\end{eqnarray}
Choosing ${\bf R}'$ along one of the coordinate axes and applying the
recurrence relation\cite[\S8.933]{GR}
\begin{eqnarray}
\cos\theta \, C_l^{\nu}(\cos\theta) &=& {1\over 2 (l + \nu)}[(l + 1) \, C_{l +
1}^{\nu} (\cos\theta) \nonumber\\
&+& (2\nu + l - 1) \, C_{l -1 }^{\nu}(\cos\theta)] ,
\end{eqnarray}
we notice that the orthogonality of the Gegenbauer polynomials (with $C_0^{\nu}
(\cos\theta) = 1$) implies that all terms except $J = 1$ are zero. From
(\ref{sphere_end}) we finally obtain the mean-square end-to-end vector for a
polymer living on the surface of a $D-1$--dimensional sphere as:
\begin{eqnarray}
\la({\bf R} - {\bf R}')^2\ra &=& 2 a^2 \left[1 - \exp\left(- {L l\over 2
a^2}\right)\right] \nonumber\\
&\simeq& \left\{
\begin{array}{ccc}
L l \; ; \; L l \ll a^2 .\\ 
2 a^2 ;\; L l \gg a^2 .
\end{array}
\right . 
\label{sphere_end_final}
\end{eqnarray}
Therefore, a characteristic area $2 a^2$ called geometrical localization area
emerges in the description of $\la({\bf R} - {\bf R}')^2\ra$, due to the
geometrical restriction of space. 

In contrast to the problem of a stiff polymer chain without any confinement,
where the chain expands due to chain stiffness, chain contracts here by
geometrical confinement. If the area of the confining sphere is small in
comparison with $L l$, then the polymer wraps many times around the sphere so
that $\la({\bf R} - {\bf R}')^2\ra$ approaches the geometrical localization
area. On the other hand if $a^2 \gg L l$, then the polymer does not sense the
curvature of the sphere so that $\la({\bf R} - {\bf R}')^2\ra = L l$. The
mean-square end-to-end distance (\ref{sphere_end_final}) normalized at $a^2$ is
plotted in Fig.~\ref{figs:fig4} as a function of the dimensionless variable ${L
l\over a^2}$. It is also to be noted that $\la({\bf R} - {\bf R}')^2\ra$ is
independent of $D$.

\subsection{\underline {The Cylinder in ${\Bbb R}^3$}}

Consider a cylinder of radius $a$ and of infinite length, with the induced
metric from ${\Bbb R}^3$ (in cylindrical coordinates, with $Oz$ along the axis
of the cylinder): $g_{\phi\phi} = a^2$; $g_{\phi z} = 0$; $g_{zz} = 1$;
$\sqrt{g} = a$.  The position vector ${\bf R}(s)$ of any bead $s$ has the
components $(a \cos\phi, a\sin\phi, z)$. The polymer dimension is $D = 2$.

The Green's function equation in the Laplace space (\ref{Green}) can be solved
directly by expanding the $\widetilde{\text{G}}$ function in the complete set
of orthogonal functions on the circle $e^{ i m(\phi - \phi ')}$ and Fourier
transforming with respect to $z - z'$, which yields
\begin{eqnarray}
\widetilde{\text{G}}({\bf R} |{\bf R}'; E ) &=& {1\over 2\pi a} \sum_{m =
-\infty}^{\infty} \! \int_{-\infty}^{+\infty} \! {d k\over 2 \pi} \,
{1\over E + {l m^2\over 4 a^2} + {l k^2\over 4}}\nonumber\\
&\times&  e^{ i m (\phi - \phi ')} \, e^{i k (z - z')} 
\end{eqnarray}
After applying an inverse Laplace transform and evaluating the integral over
$k$, the spectral expansion of $\text{G}({\bf R}| {\bf R}' ; L)$ is found to
be
\begin{eqnarray}
\text{G}({\bf R} | {\bf R}'; L) & = & {1\over 2 \pi a} \, \sqrt{1\over \pi L l}
\, e^{- {(z - z')^2\over L l}} \nonumber\\
&\times& \sum_{m = -\infty}^{\infty} \! e^{i m (\phi -
\phi ')} \, e^{- {L l m^2\over 4 a^2}}.
\label{green_cyl}
\end{eqnarray}
An equivalent form of the expression above can be obtained by using the
relation
\begin{equation}
\exp(- \lambda x^2) = {1\over \sqrt{2 \pi}}\, \int_{-\infty}^{+\infty} \! d\xi
\, e^{ - {1\over 2} \xi^2 + i \sqrt{2 \lambda} \, \xi \, x} \;\;\; (\lambda >
0) , \label{Straton}
\end{equation}
and the Poisson formula
\begin{equation}
\sum_{m = -\infty}^{+\infty} e^{2 i \pi \alpha m} = \sum_{ n =
-\infty}^{+\infty} \delta (\alpha - n), \;\;\; (m,n \in {\Bbb Z}).
\label{Poisson}
\end{equation}
Applying these transformations in Eq.~(\ref{green_cyl}), we arrive at the the
{\sf winding number} expansion
\begin{equation}
\text{G}({\bf R} | {\bf R}'; L) = {1\over \pi L l} \, e^{- {(z - z')^2\over L
l}} \sum_{n = -\infty}^{\infty} \! e^{- {a^2 (\phi - \phi ' + 2\pi n)
\over L l}}, 
\label{green_cyl_wind}
\end{equation}
where $n$ (a topological term) is the {\sf winding number} that counts how many
times the chain winds around the cylinder.  A term in the sum represents now
the probability that a given chain starting at ${\bf R}$ and ending at ${\bf
R}'$ will wind around the cylinder $n$ times (in $L/l$ steps). This expansion
can be used to investigate the statistics of the winding number (e.g. the
mean-square winding number $\la N^2\ra$). 

To calculate the mean-square end-to-end vector, we insert $({\bf R} - {\bf
R}')^2 = (z - z')^2 + {2 a^2 [1 - \cos(\phi - \phi ')]}$ and the distribution
function (\ref{green_cyl}) in (\ref{end_to_end}). Evaluating the integrals, we
get 
\begin{eqnarray}
\la({\bf R} - {\bf R}')^2\ra &=& {L l\over 2} + 2 a^2 \left[1 - e^{- {L l\over
4 a^2}}\right] \nonumber\\
&\simeq& \left\{
\begin{array}{ccc}
L l \,;\; L l \ll a^2 .\\ {L l\over 2} ;\; L l \gg a^2 .
\end{array}\right. 
\label{cyl_end_final}
\end{eqnarray}
As expected, the polymer remains almost Gaussian (a prefactor of $1/2$ instead
of 1) along the $\mathit{Oz}$ direction. The influence of the geometry appears
through the radius $a$, related to the mean curvature of the cylinder. Once
again, the polymer becomes {\sf geometrically localized}.  The mean-square
end-to-end distance (normalized at $a^2$) is represented in Fig.~\ref{figs:fig4} as
a function of the dimensionless variable ${L l\over a^2}$.

\subsection{\underline{The Cone in ${\Bbb R}^3$}}

Now let us consider the surface of cone with the vertex at the origin $O$,
centered about the $Oz$ axis and with the opening angle $2 \alpha$. In
cylindrical coordinates, the position vector ${\bf R}(s)$ of any point $s$
along the chain is ${\bf R}(s) = (\rho(s) \cos\phi(s), \rho(s) \sin\phi(s),
\rho(s) \cot\alpha)$ and the induced metric is: $g_{\rho\rho} = 1 +
\cot^2\alpha$; $g_{\rho\phi} = 0$; $g_{\phi\phi} = \rho^2$; $\sqrt{g} = \rho
\sqrt{1 + \cot^2\alpha}$. As in the previous problem, the dimension of the
polymer is $D = 2$.

To compute the probability distribution $\text{G}({\bf R} | {\bf R}'; L)$ of
the end-to-end vector we start with the Green's function equation (\ref{Green})
obeyed by the Laplace transform $\widetilde{\text{G}}$:
\begin{eqnarray}
&& \left( E  -{l\over 4 \rho } \sin^2\alpha {\partial\over
\partial\rho} \rho 
{\partial\over \partial\rho} - {l\over 4 \rho^2} {\partial^2\over
\partial\phi^2} \right) \widetilde{\text{G}}({\bf R} | {\bf R}'; E) \nonumber\\
& &  = {\sin\alpha \over \rho} \, \delta (\rho - \rho ') \, \delta (\phi -
\phi') 
\label{diffusion_cone}
\end{eqnarray}
We solve this equation in a standard manner \cite{COURANT}. First we eliminate
the azimuthal dependence of $\widetilde{\text{G}}$ by inserting the expansion
\begin{eqnarray}
\widetilde{\text{G}}({\bf R} | {\bf R}'; E) &=& {1\over 2\pi} \sum_{m  =
-\infty}^{+\infty} \tilde{\text{g}}_m (\rho | \rho '; E) \nonumber\\
&\times & \exp[i m (\phi - \phi ')],
\label{Laplace_cone}
\end{eqnarray}
in (\ref{diffusion_cone}), which gives
\begin{eqnarray}
&&\left[{1\over x} \, {\partial\over \partial x} x {\partial\over
\partial x}  - (1 + {\nu^2 \over x^2})\right] \tilde{\text{g}}_{\nu}
(x | x'; E) \nonumber\\
&& =  - {4 \over x l \sin\alpha } \, \delta (x - x') ,
\end{eqnarray}
where we introduced the new variable $x = k\rho$, with $k^2 =
{4 E \over l \sin^2 \alpha}$ and $\nu^2 = {m^2\over \sin^2\alpha}$. The 
solutions of the homogeneous equation are the modified Bessel functions
$\text{I}_{|\nu|}(x)$ and $\text{K}_{|\nu|}(x)$. Then, the regularity of
$\tilde{\text{g}}_{\nu} (x | x'; E)$ at $x = 0$, $ x 
\to\infty $, the continuity at $ x = x'$ and the jump in the first derivative
impose the solution
\begin{equation}
\tilde{\text{g}}_{\nu} (\rho | \rho'; E) = {4\over l \sin\alpha}\,
\text{I}_{|\nu|}(k \rho_<) \, \text{K}_{|\nu|}(k \rho_>) .  
\end{equation}
Here $\rho_<$ and $\rho_>$ denote the smaller and the larger, respectively,
between the variables $\rho$ and $\rho'$; we have also used the fact that the
Wronskian W of the modified Bessel functions is ${\text{W}[\text{I}_{\nu}, \,
\text{K}_{\nu}](x) = - {1\over x}}$. Inserting the solution found back in
Eq.~(\ref{Laplace_cone}) and performing the inverse Laplace transform using the
formula \cite[\S 5.16.56]{MOTRANS}
\begin{eqnarray}
{\cal L}^{-1} \left\{ \text{K}_{\nu}[(\sqrt{a} + \sqrt{b})\sqrt{E}] \,
\text{I}_{\nu} [(\sqrt{a} - \sqrt{b})\sqrt{E}]\right\} & & = \\
{1\over 2 L} \, e^{- {a + b\over 2 L}} \,  \text{I}_{\nu}({a - b\over 2 L})
\;\;\; (\Re a, \Re b > 0) , \nonumber  
\end{eqnarray}
we arrive at the exact spectral expansion of $\text{G}({\bf R} | {\bf R}'; L)$
for a polymer on a cone:
\begin{eqnarray}
\text{G}({\bf R} | {\bf R}'; L) & = & {1\over \pi L l \sin\alpha}
\sum_{m=-\infty}^{+\infty} \text{I}_{\left|{m\over \sin\alpha}\right|}
(2 \rho\rho '/ (L l \sin^2\alpha)) \nonumber\\  
& \times & \exp\left[- {\rho^2 + \rho '^2\over L l
\sin^2\alpha}\right] \, \exp[i m (\phi - \phi')],
\end{eqnarray}
which is properly normalized to one ($\int\! d{\bf R} \, \text{G}({\bf R} |
{\bf R}'; L) = 1$). For $\alpha = {\pi\over 2}$ we recover the expression in
polar coordinates of the propagator of a Gaussian polymer chain in the
plane \cite{KHANDEKAR,WIEGEL}.

As the total partition sum gives the area of the cone, we
calculate the end-to-end distance by fixing ${\bf R}'$ and integrating only
over ${\bf R}$. Using cylindrical coordinates, with $({\bf R} - {\bf R}')^2 =
(\rho^2 + \rho'^2)/\sin^2\alpha + 2\rho\rho' [1 - \cos(\phi-\phi')]$ and
applying various formulae \cite{GR} involving definite integrals of Bessel
functions, we obtain
\begin{eqnarray}
&&\la({\bf R} - {\bf R}')^2\ra= L l \, \left\{ 1 + 2 \, t^2 \,
\csc^2\alpha \right. \nonumber\\
&& - \sqrt\pi \csc\alpha \,(1 - \sin^2\alpha)\,  _1\text{F}_1
(-1/2, 1; -t^2 \csc^2\alpha)\nonumber\\
& & - \left. 2 \,{\Gamma (3/2 + 1/2 \csc\alpha)\over \Gamma(1 + \csc\alpha)} \,
(\sin\alpha)^{(1 - \csc\alpha)} \, t^{(1 + \csc\alpha)}
\right. \label{cone_end_final}\\
& &\times \left. _1\text{F}_1 [(\csc\alpha - 1)/2, 1 + \csc\alpha ; - t^2 \,
\csc^2\alpha] \right\},  \nonumber
\end{eqnarray}
where $t = \rho'/ (L l)^{1/2}$ and $_1\text{F}_1$ is the confluent
hypergeometric function. 

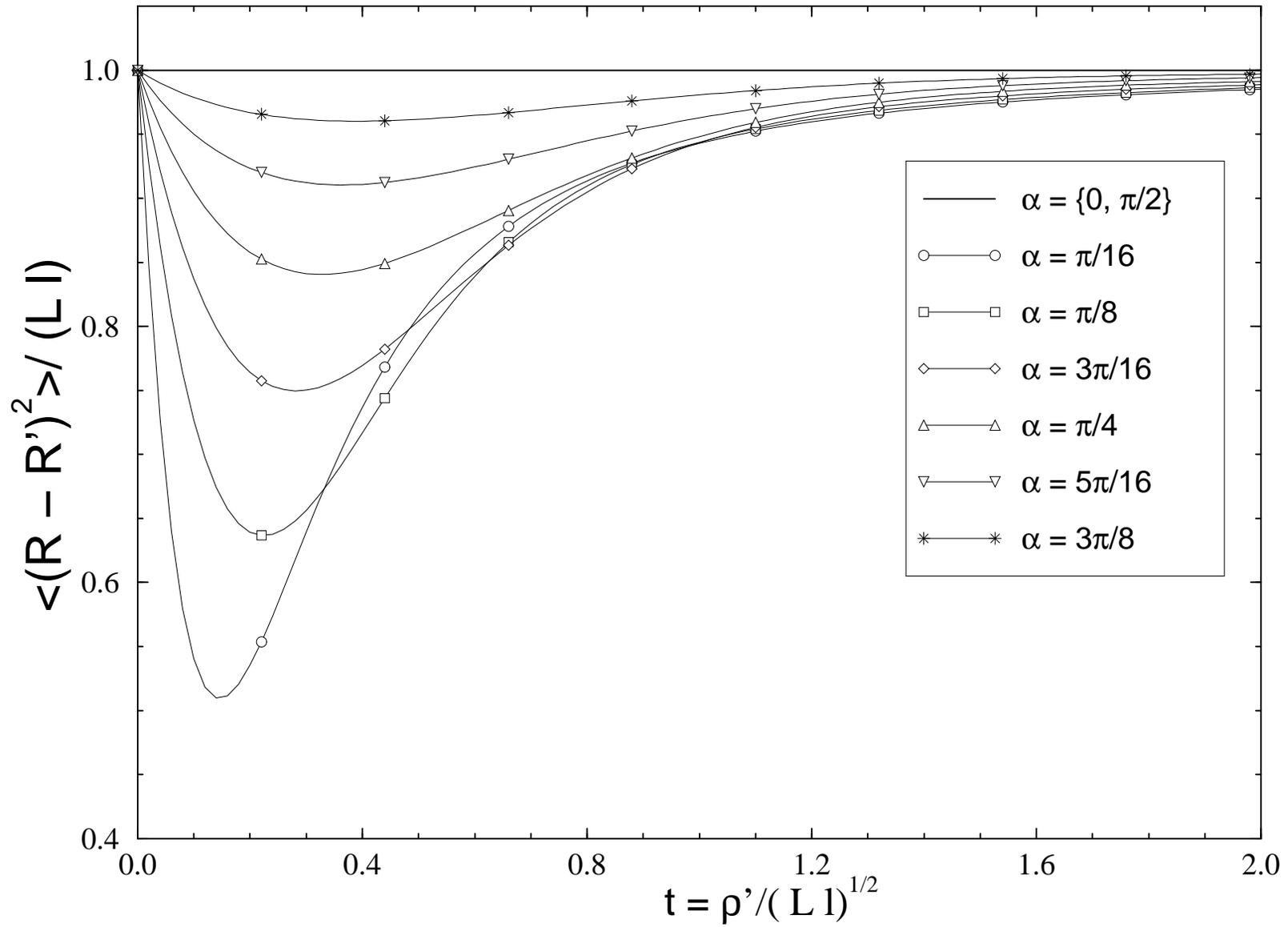
\begin{figure}
\caption{The mean-square end-to-end distance $\la ({\bf R} - {\bf R}')^2 \ra$
normalized to $L l$ for a Gaussian polymer chain embedded on a cone, as a
function of $t = {\rho '\over\protect\sqrt{L l}}$ and $\alpha$.}
\label{figs:cone}
\end{figure}

Rather surprisingly, the mean-square distance depends on the position of the
chain through the parameter $t$. To interpret this finding, we calculate the
limiting values of $\la({\bf R} - {\bf R}')^2\ra$ for $t\to\infty$ (polymer far
on the conical surface or $\alpha\to 0$), $t\to 0$ (polymer with one end
attached to the cone vertex) and $\alpha = {\pi\over 2}$ (polymer on a $2$--$D$
plane surface). One finds that
\begin{equation}
\la({\bf R} - {\bf R}')^2\ra = L l \;\;\text{for}\;\; \{t = 0;\, t
\to\infty;\,\alpha = {\pi\over 2}\}, 
\end{equation}
which is just the Gaussian result in all cases. Physically, when the polymer
chain is either fixed at the origin ($t = 0$) or far away from it
($t\to\infty$), there are no other preferred points on the cone and $\la({\bf
R} - {\bf R}')^2\ra$ has the planar value (we should also remark that a cone is
geometrically flat). In the crossover region, the singular character of the
cone vertex becomes dominant and induces the {\sf contraction} of the chain
(possibly by winding about the origin). This behavior is illustrated in
Fig.~\ref{figs:cone} and Fig.~\ref{figs:cone2}, where the mean-square distance
$\la({\bf R} - {\bf R}')^2\ra$ normalized at $L l$ is plotted as a function of
the $t$ parameter and the angle $\alpha$.

\subsection{\underline{The Torus in ${\Bbb R}^3$}}

Consider a torus embedded in ${\Bbb R}^3$, with the circular cross-section of
radius $a$. The axial circle containing the centers of the circular sections
has the radius $b > a$ (see Fig.~\ref{figs:geometry}). The torus is centered
about the $Oz$ axis and symmetric with respect to the $xOy$ plane. We introduce
the toroidal coordinates $\{\eta, \theta,\phi\}$, \cite[\S 2.13]{TOR} where
$\phi$ is the usual azimuthal angle---in cylindrical coordinates---about $Oz$:
\begin{eqnarray}
&&x = {c \sinh\eta \cos\phi\over \cosh\eta - \cos\theta} ;\;\; 
y = {c \sinh\eta \sin\phi\over \cosh\eta - \cos\theta} ;\nonumber\\
&& z = {c \sin\theta \over \cosh\eta - \cos\theta} ,
\end{eqnarray}
where $0 \le \eta < \infty$, $0 \le \theta < 2 \pi$ and $0 \le \phi < 2
\pi$. The surface of the torus corresponds to a fixed value $\eta_0$ of the
coordinate $\eta$. Then we have the following geometrical relations:
\begin{equation}
b = c \coth\eta_0 ; \;\; a = {c \over \sinh\eta_0}; \;\; c^2   = b^2 - a^2 .
\end{equation}

The induced metric tensor is $g_{\theta\theta} = {c^2 \over (\cosh\eta_0 -
\cos\theta)^2}$; $g_{\phi\phi} = {c^2 \sinh^2\eta_0 \over (\cosh\eta_0 -
\cos\theta)^2}$; $g_{\theta\phi} = g_{\phi\theta} = 0$; $\sqrt{g} = {c^2
\sinh\eta_0 \over (\cosh\eta_0 - \cos\theta)^2}$. We recall that the length
interval on the surface is $ds^2 = g_{\theta\theta} \, d\theta^2 +
g_{\phi\phi} \, d\phi^2$ and the volume element is $d^2\rvec = \sqrt{g} \,
d\theta\,d\phi$. 

The calculation of the propagator of a Gaussian chain on a torus in ${\Bbb
R}^3$ (which is equivalent to solving the problem of a free quantum particle
moving on the same surface) cannot be done in closed form, as involves finding
the eigenvalues of a Hill-type equation. Still, using the path-integral
formalism and making certain ``sloppy'' approximations, we were able to derive
an approximate form of $\text{G}({\bf R} | {\bf R}'; L)$ that behaves
physically correct when computing moments of the distribution of ${\bf R} -
{\bf R}'$, which are the experimentally relevant quantities we are interested
in. This approach will be justified {\it a posteriori} by showing that one
recovers the usual flat surface case and the expected asymptotic regimes
obtained for spheres and cylinders.

We start be recalling the path-representation (\ref{path_int}) written in a
compact form as
\begin{equation}
\text{G}({\bf R} | {\bf R}'; L) = \int_{\rvec_0 = {\bf R}'}^{\rvec_N = {\bf R}}
{\cal D}[\rvec]  \, \exp\left( - \sum_{j=1}^N \text{S}_j \right) ,
\label{path_int_torus}
\end{equation}
where the integration measure is given explicitly by
\begin{equation}
{\cal D}[\rvec]  = \lim_{\stackrel{\scriptstyle
N\to\infty}{\stackrel{\scriptstyle \delta s_j\to 0}{\sum_j\!\delta s_j = L}}}
\prod_{j = 1}^N \! {1\over \pi l \delta s_j} \, \prod_{j = 1}^{N-1}\!d^2\rvec_j
\label{meas_torus}
\end{equation}
and $\text{S}_j$ is the {\sf short-length} polymer action
\begin{equation}
\text{S}_j  = {(\rvec_j - \rvec_{j-1})^2 \over l \delta s_j} .
\label{action_torus}
\end{equation}
In the previous expressions, $\delta s_j = s_j - s_{j - 1}$ (with $j =
\overline{1,N}$, $s_0 = 0$, $s_N = L$) and $\rvec_j$ is the position vector
about the origin of the $j$-th element of the discretized chain.

In order to evaluate the path-integral, due to the mentioned equivalence
between the statistical mechanics of polymers and quantum mechanics, we will
apply the method used to calculate the propagator of a free particle
moving on a circle and in the presence of a ring-shaped
defect \cite{KLEINERT,WIEGEL,RAMOS}.

We start with the action element $\text{S}_j$. Using the length interval on the
torus, we write  $\text{S}_j$ in a symmetrized form
\begin{eqnarray}
\text{S}_j &=& {c^2 \over l} {1 \over \delta s_j (\cosh\eta_0 -
\cos\theta_j)(\cosh\eta_0 - \cos\theta_{j - 1})}\nonumber\\
&\times & [(\Delta\theta_j)^2 + \sinh^2\eta_0 \, (\Delta\phi_j)^2] ,
\label{action_torus_SJ}
\end{eqnarray}
where $\{\theta_j = \theta(s_j), \phi_j = \phi(s_j)\}$ are the toroidal
coordinates of the element located at $s_j$ along the chain, ${\Delta\theta_j =
\theta_j - \theta_{j - 1}}$ and $\Delta\phi_j = \phi_j - \phi_{j - 1}$.

Next we apply a local rescaling \cite{KHANDEKAR,MeV}, which reads
\begin{equation}
\sigma_j = \delta s_j \, (\cosh\eta_0 - \cos\theta_j)(\cosh\eta_0 -
\cos\theta_{j - 1})
\label{scaling}
\end{equation}
and which satisfies the global scaling relation:
\begin{equation}
\sigma = \sum_{j=1}^N \sigma_j = L (\cosh\eta_0 - \cos\theta) (\cosh\eta_0 -
\cos\theta ') .
\label{global}
\end{equation}
This constraint is nontrivial, as it must be compatible with the local scaling
(\ref{scaling}).  That it gives the correct answer when applied in the path
integral and when one uses an anisometric discretization (unequal length
intervals $\delta s_j$), was discussed by Inomata \cite{MeV}. Basically, it
amounts to the prescription \cite{FEYN} of ignoring terms of
$O(\delta s_j^{1 + \epsilon})$ in the path-integral, if $\epsilon > 0$.  Then,
the discretized functional measure becomes:
\begin{eqnarray}
{\cal D}[\rvec] & = & (\cosh\eta_0 - \cos\theta) (\cosh\eta_0 - \cos\theta ')\,
\nonumber\\ 
& \times & \prod_{j = 1}^N \! {1\over \sigma_j} \, \prod_{j = 1}^{N-1} \!
d\theta_j \, d\phi_j \, c^2 \sinh\eta_0 .
\end{eqnarray}
We apply now the length rescaling in the action $\text{S}_j$ given by
(\ref{action_torus_SJ}) directly, by dropping boldly any terms coming from a
rigorous expansion of $\Delta\theta_j$ and $\Delta\phi_j$ about the new length
variable $\sigma_j$. As mentioned, this will be justified by the final
results. Combining then with the integration measure, we get the approximate
discretized propagator $\text{G}_N$
\begin{eqnarray}
\text{G}_N({\bf R}& | & {\bf R}'; \sigma) \approx {1\over c^2} (\cosh\eta_0 -
\cos\theta) (\cosh\eta_0 - \cos\theta ') \nonumber\\
& \times &  \prod_{j = 1}^N \! {c^2 \over \pi l \sigma_j} \, \int_0^{2\pi} \!
\prod_{j = 1}^{N-1}\! d\theta_j \, \exp\left[ - {c^2\over l} \sum_{j = 1}^N
{(\Delta\theta_j)^2 \over \sigma_j}\right]  \nonumber\\
& \times & \int_0^{2\pi} \!  \prod_{j = 1}^{N-1}\! d\phi_j \,
\sinh\eta_0\label{green_torus}  \\
&\times & \exp\left[ - {c^2 \sinh^2\eta_0 \over l} \sum_{j = 1}^N
{(\Delta\phi_j)^2 . \over \sigma_j}\right] \nonumber
\end{eqnarray}
Observing that the integrals over the angular coordinates are similar to the
path integral for a free particle moving on a circle (e.g. see Ref.\
[\onlinecite{KLEINERT,KHANDEKAR}]), we have the formula
\begin{eqnarray}
\prod_{j = 1}^N \! \left({\alpha \over \pi \sigma_j}\right)^{1/2} & & 
\int_0^{2\pi} \! \prod_{j = 1}^{N-1} \! d\psi_j \, e^{- \alpha \sum_{j = 1}^N {
(\Delta\psi_j)^2 \over \sigma_j}} \nonumber\\
= & & \sqrt{\alpha \over \pi \sigma} \sum_{n = -\infty}^{\infty} e^{-
{\alpha\over \sigma } (\psi - \psi ' + 2 \pi n)^2}.
\label{formula_torus}
\end{eqnarray}
Here $\alpha$ is an arbitrary positive constant and $\psi_j = \psi (\sigma_j)$
is the discretized angular variable on the circle, with $\psi'$ and $\psi$
denoting the initial and the final positions. Equivalently, recognizing that
the angular path integral contains trajectories that wind a different number of
times around the origin, one can use the covering space mapping
\cite{KLEINERT,KHANDEKAR}. 

Returning to Eq.~(\ref{green_torus}) of the approximate discretized propagator
$\text{G}_N$, integrating over the angles using formula (\ref{formula_torus}),
and replacing $\sigma$ with its global value (\ref{global}), we finally
obtain---up to a normalization factor---the approximate distribution function
of the end-to-end distance for a polymer chain on a curved torus in ${\Bbb
R}^3$, expressed as a {\sf winding number} expansion:
\begin{eqnarray}
\text{G}({\bf R} &|& {\bf R}' ; L) \approx  {1\over \pi L l} \,
\sum_{n=-\infty}^{\infty} \sum_{m = -\infty}^{\infty} \exp\left\{ - {c^2\over L
l} \right. \nonumber\\
&\times& {1\over (\cosh\eta_0 - \cos\theta)(\cosh\eta_0 -
\cos\theta')}\label{green_torus_wind}\\ 
&\times & \left. [(\theta - \theta' + 2\pi n)^2 + (\sinh^2 \eta_0) (\phi -
\phi' + 2\pi m)^2] \right\}\nonumber
\end{eqnarray}
Another compact and illuminating form is obtained by employing the
transformations (\ref{Straton}) and (\ref{Poisson}), which yield the {\sf
angular momentum} expansion 
\begin{eqnarray} 
\text{G}({\bf R} &|& {\bf R}'; L)\approx  {1\over 4 \pi^2} {(\cosh\eta_0 -
\cos\theta)(\cosh\eta_0 - \cos\theta')\over c^2 \sinh\eta_0} \, \nonumber\\
& \times & \Theta_3[(\theta - \theta')/2, i\sigma(\theta,\theta') l/(4 \pi
c^2)]\, \label{green_torus_theta}\\
&\times& \Theta_3[(\phi - \phi')/2, i\sigma (\theta,\theta') l/(4 \pi c^2
\sinh^2\eta_0)] , \nonumber
\end{eqnarray}
where we recall that $\sigma(\theta, \theta')$ $=$ $L (\cosh\eta_0$ $-
\cos\theta)$ $(\cosh\eta_0$ $- \cos\theta')$, $c^2 = b^2 - a^2$ and
$\cosh\eta_0 = {b\over a}$. $\Theta_3$ is the theta function defined as
\cite{GR} 
\begin{equation}
\Theta_3 (z,\tau) = \sum_{n=-\infty}^{\infty} e^{i\pi \tau \, n^2} \, e^{2 i z
n} \;\;\; (\Im\tau > 0) .
\label{theta}
\end{equation}
First, we stress that this is not the exact Green function of the heat equation
on the curved torus. Still, as necessary, the expression is symmetric in ${\bf
R},{\bf R}'$, translationally invariant with respect to the azimuthal angle
$\phi$, and in the limit $\eta_0\to\infty\equiv a\to 0$ reduces properly to the
probability distribution of the end-to-end vector for a polymer on a circle
(see the angular part in Eq.\ (\ref{green_cyl_wind})). Also, it obeys the
initial condition
\begin{equation}
\text{G}({\bf R} | {\bf R}'; 0) = \delta^{(2)}({\bf R} - {\bf R}'),
\end{equation}
and its trace has, at least numerically, the correct limit \cite{KEAN} when ${L
l\over a^2}\to 0$:
\begin{eqnarray}
\text{Z}(x \to 0, t) &=& \int\!\!\!\int\! d^2{\bf R} \, \text{G}({\bf R},L |
{\bf R},0) \label{trace}\\
&\simeq& 4 \pi {t \over x} + {\pi^3\over 60} \,
x \, {t\, (t - \sqrt{t^2 - 1})\over \sqrt{t^2 - 1}} + o(x^2),\nonumber
\end{eqnarray}
where we introduced the dimensionless variables:
\begin{equation}
t = \cosh\eta_0 = {b\over a} ; \;\;\; x = {L l \over a^2}.
\label{dimless_torus}
\end{equation}
A closed formula for $\la({\bf R} - {\bf R}')^2\ra$ cannot be readily obtained
and a numerical calculation is required. To begin with, we write the distance
(in ${\Bbb R}^3$) between the ends of the polymer chain:
\begin{eqnarray} 
({\bf R} - {\bf R}')^2 & = & {2 c^2\over (\cosh\eta_0 -
\cos\theta)(\cosh\eta_0 - \cos\theta')} \, [\cosh^2\eta_0 \nonumber\\
& - & \sinh^2\eta_0 \cos(\phi - \phi') - \cos(\theta - \theta')].
\label{dist_torus}
\end{eqnarray}
Using the Green function (\ref{green_torus_theta}) in the formula for the
mean-square end-to-end distance (\ref{end_to_end}) with the proper volume
element on the surface of the torus, evaluating the integrals over the
azimuthal angles $\phi$ and $\phi'$ by applying the definition (\ref{theta}) of
the theta function and the formula:
\begin{equation}
\int\!\!\!\int_0^{2\pi}\! d\phi\,d\phi' \: \cos(\phi -\phi') \, e^{i m(\phi -
\phi')} = 2 \pi^2 (\delta_{-1m} + \delta_{+1m}),
\end{equation}
and expressing all quantities in terms of the dimensionless variables from
Eq.~(\ref{dimless_torus}), we eventually obtain the mean-square end-to-end
distance as
\begin{eqnarray}
&&{\la ({\bf R} -  {\bf R}')^2 \ra \over a^2} = {2 (t^2 - 1)\over N(x,t)}
\, \int\!\!\!\int_0^{2 \pi}\! d\theta \, d\theta' \, \nonumber\\
& & \;\; \times {t^2 + (1 - t^2) e^{-{x (t - \cos\theta)(t - \cos\theta')\over
4 (t^2 - 1)^2}} - \cos(\theta - \theta') \over (t - \cos\theta)^2(t -
\cos\theta')^2}\label{dist_torus_final}\\
& & \;\; \times \Theta_3 \left[{\theta - \theta'\over 2}, {i x (t -
\cos\theta)(t - \cos\theta') \over 4 \pi (t^2 - 1)}\right]  , \nonumber
\end{eqnarray}
where $N(x,t)$ is given by
\begin{eqnarray}
N(x,t) & = & \int\!\!\!\int_0^{2 \pi}\! d\theta \, d\theta' \, {1\over (t -
\cos\theta)(t - \cos\theta')} \nonumber\\
& & \times \Theta_3 \left[{\theta - \theta'\over 2}, {i x (t - \cos\theta)(t -
\cos\theta') \over 4 \pi (t^2 - 1)}\right] .
\end{eqnarray}
The partition function $Z(x,t)$ of the polymer on the torus is:
\begin{equation}
Z(x,t) = a^2 \, (t^2 - 1)^{3\over 2} N(x,t).
\end{equation}
This reduces properly to the area $4 \pi^2 a b$ of the torus when
$x\gg 1$, $t\gg 1$ ($\Theta(z,\tau) \simeq 1$):
\begin{equation}
Z(x,t) \simeq 4 \pi^2 \,a b\, \sqrt{1 - {1\over t^2}} \simeq 4 \pi^2 a b 
\end{equation}
Physically, for different values of $x$ and $t$, three regimes are expected:
\begin{mathletters}
\label{torus_asymp}
\begin{eqnarray}
a) & & \;\;\; L l \ll a^2 < b^2 \iff {x\over t^2} < x \ll 1  ,\\
b) & & \;\;\; a^2 \ll L l \ll b^2 \iff {x\over t^2} \ll 1 \ll x , 
\label{torus_asymp_cyl}\\
c) & & \;\;\; a^2 < b^2 \ll L l \iff 1 \ll {x\over t^2} < x ,
\label{torus_asymp_sphere}
\end{eqnarray}
\end{mathletters}
where in the first case we should recover the solution for a polymer in the
plane, in the second we must obtain the result for a polymer on a cylinder and
in the third, the mean square end-to-end distance should reach a constant value
(corresponding to a large winding number about both the $Oz$ axis and the axial
circle of radius $b$). 

These regimes are manifest in Fig.~\ref{figs:fig4}, where the results for the torus
are plotted with thin lines, for $t = \{2;4;6;8;10\}$. 

Initially (the first regime in Eq.~(\ref{torus_asymp})), the polymer is too
small to explore the geometry of the surface, and we recover (numerically) the
case of a polymer in a plane. This is also the behavior found for the other
surfaces. 

As the length of the chain increases, it starts winding along the circle of
radius $a$ and then along the axial circle, about the $Oz$ axis. When $t$
increases, the behavior is similar---for a certain range of $x$ values---to
that of a chain on a cylinder: the mean-square end-to-end distance is linear in
$L l/2$.  We can check analytically that the approximate propagator from
Eq.~(\ref{green_torus_theta}) produces the correct behavior. Because $a^2 \ll L
l \ll b^2$, the term $n = 0$ dominates in the theta function in
Eq.~(\ref{dist_torus_final}) ($\Theta_3(z,\tau) \approx 1$) and the exponential
coefficient of $1 - t^2$ can be expanded, so one obtains explicitly
\begin{equation}
\la({\bf R} - {\bf R}')^2\ra \simeq 2 a^2 + {L l\over 2} \;\;\;\; (a^2 \ll L l
\ll b^2) ,
\end{equation}
which is just the limit for the polymer on the cylinder of radius $a$ from
Eq.~(\ref{cyl_end_final}). 

Eventually, at any given $t$ and large enough $x$, $\la({\bf R} - {\bf
R}')^2\ra$ departs from the behavior of a polymer on a cylinder and approaches
a constant value, which is a function only of the geometrical parameters $a$
and $b$. This limiting value can be explicitly calculated. Now we have $a^2 <
b^2 \ll L l$, thus once again $\Theta_3(z,\tau) \approx 1$ but the exponential
coefficient of $1 - t^2$ in the formula for distance (\ref{dist_torus_final})
is zero, which gives the limit:
\begin{equation}
\la ({\bf R} - {\bf R}')^2 \ra \approx 2 \, (a^2 + b^2) \;\;\; (a^2 < b^2 \ll L
l) . 
\end{equation}
If either $a$ or $b$ radii becomes $0$, one recovers the asymptotic
limits of the end-to-end distance for a polymer on a circle or on a sphere,
respectively, as found in Eq.~(\ref{sphere_end_final}). 

Although the asymptotic behavior is correctly recovered, we expect our results
in the crossover region of intermediate values of $t$ and $x$ to be only
approximate. 

\section{Conclusions}

Due to its relevance to studies of the behavior of polymers confined at
interfaces, we have considered here the problem of a linear, Gaussian polymer
chain embedded in the following surfaces: the $\text{S}^{D-1}$ {\sf sphere}
in $D\/$ dimensions, the {\sf cylinder}, the {\sf cone} and the {\sf curved
torus} in ${\Bbb R}^3$.

We obtained closed formulas for the probability distribution function
$\text{\bf G}({\bf R} | {\bf R}' ; L)$ of the end-to-end vector of the chain
for the sphere, the cylinder and the cone and an approximate propagator for the
torus.  We calculated, analytically in the case of the sphere, the cylinder and
the cone, and numerically for the torus, the {\sf mean-square end-to-end
distance} (in the embedding space) of the chain. As such, at least in the
limiting regimes previously described, the results are also valid for a free
quantum particle on a torus if one uses the mapping $\{L\to t - t'; {2\over
l}\to {m\over i\hbar}\}$ (where $t$ and $t'$ represent the final and initial
time coordinates of the particle).

Our calculations demonstrate the role played by the geometry ({\sf curvature}
and {\sf shape}) of the interface in controlling the size of the chain.

As the size of the confining surface decreases, the polymer size is determined
by the parameters of the confining surface instead of chain length. The
crossover between the free Gaussian chain limit and the confined limit is
controlled by a characteristic geometrical localization area $A$. $A$ is
proportional to $a^2$ for the cases of spheres and cylinders, and there are two
characteristic areas proportional to $a^2$ and $b^2$ for the case of curved
torus. For the cone, the localization area is variable and is determined by the
position of the singular point (the vertex) about the chain and by the angle
$\alpha$. Explicit closed formulas are derived for the dependence of the
mean-square end-to-end distance of a Gaussian chain on the geometrical
parameters of the confining surface. The formulas for spheres and cylinders are
simple (Eqs.~(\ref{sphere_end_final},\ref{cyl_end_final})), describing the
crossover between free polymer regime and confined polymer regime. For the case
of a cone, the final expression is more complicated, but still exact. For a
torus, there are three asymptotic regimes as outlined by
Eq.~(\ref{torus_asymp}), and the crossover behavior is obtained numerically
from Eq.~(\ref{dist_torus_final}).
  
A number of open problems come into attention: the calculation of the
mean-square {\sf geodesic} end-to-end distance (as measured on the surface and
not in the embedding space), the statistics of the winding numbers (mainly for
the polymer on the torus case), the influence of the topological defects
existing on the surface, the presence of excluded volume and other types of
potential interactions. We hope to address some of these issues in the near
future.

\acknowledgments 

Acknowledgment is made to the CUMIRP and Materials Research Science and
Engineering Center at the University of Massachusetts, and the NSF Grant DMR
9625485. The authors are grateful to E.~Cattani, J.~Douglas, J.~Machta and
F.~Pedit for useful discussions.

\end{multicols}

\begin{figure}
\caption{The mean-square end-to-end distance $\la ({\bf R} - {\bf R}')^2 \ra$
normalized to $L l$ for a Gaussian polymer chain embedded on a cone, as a
function of $t = {\rho '\over\protect\sqrt{L l}}$ and $\alpha$: 3--$D$
representation.} 
\label{figs:cone2}
\end{figure}

\begin{figure}
\caption{The mean-square end-to-end distance $\la ({\bf R} - {\bf R}')^2 \ra$
normalized to $a^2$ for a Gaussian polymer chain embedded on different
surfaces, as a function of ${L l \over a^2}$ and $t = {b\over a}$. The
symbols joined by thin lines represent the data for the polymer on the surface
of the curved torus in ${\Bbb R}^3$: {\sf thick solid line}, the cylinder in
${\Bbb R}^3$; $\circ$---$\circ$~torus, $t = 10$; $\Box$---$\Box$~torus, $t =
8$; $\Diamond$---$\Diamond$~torus, $t = 6$; $\triangle$---$\triangle$~torus, $t
= 4$; $\triangledown$---$\triangledown$~torus, $t = 2$; {\sf thick dotted
line}, the sphere $\text{S}^{D-1}$ in ${\Bbb R}^D$.}
\label{figs:fig4}
\end{figure}


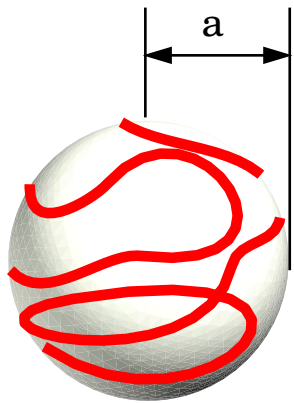
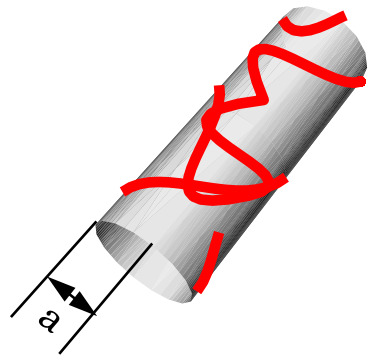
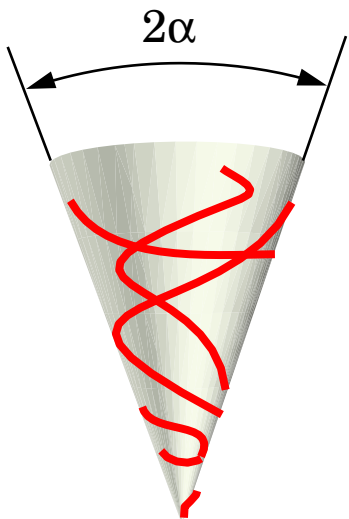
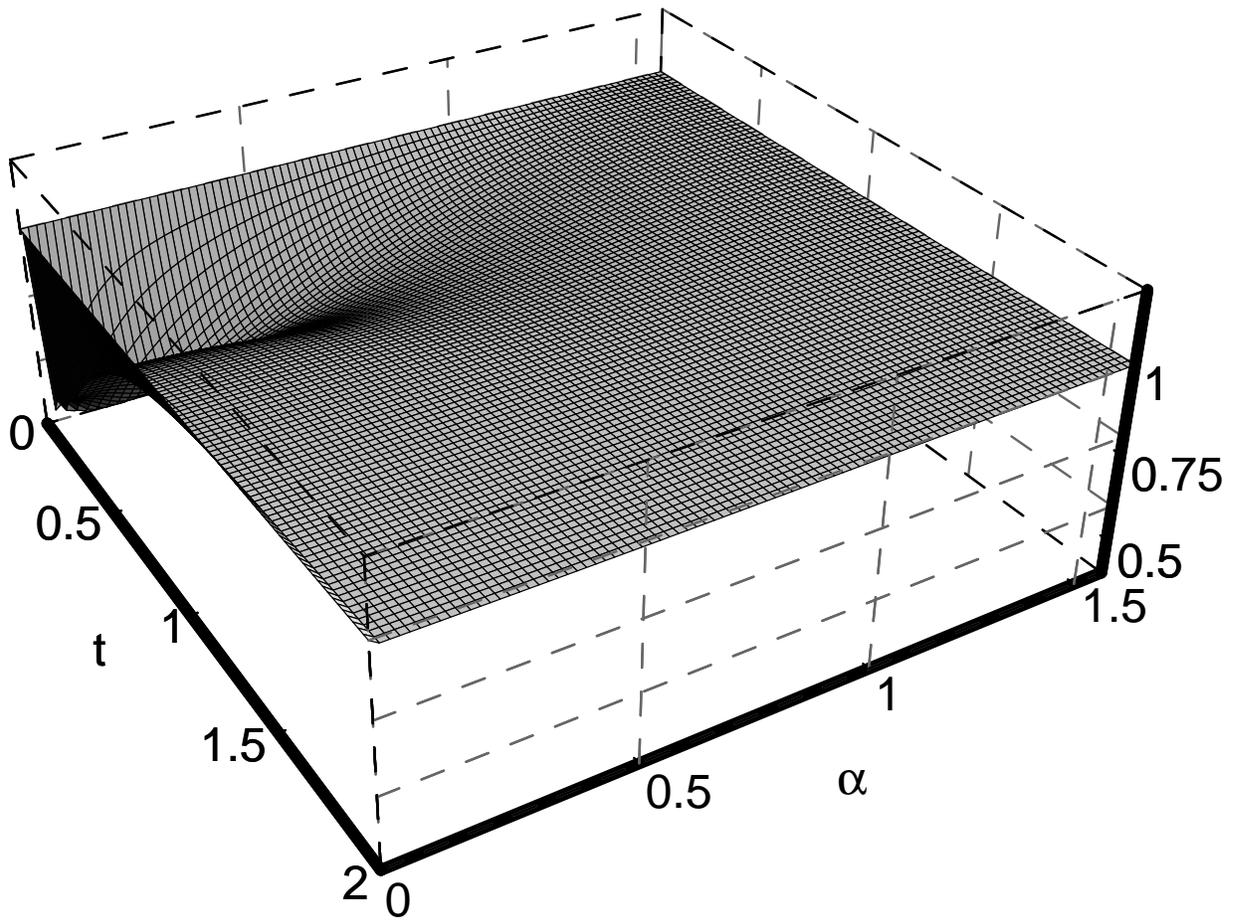
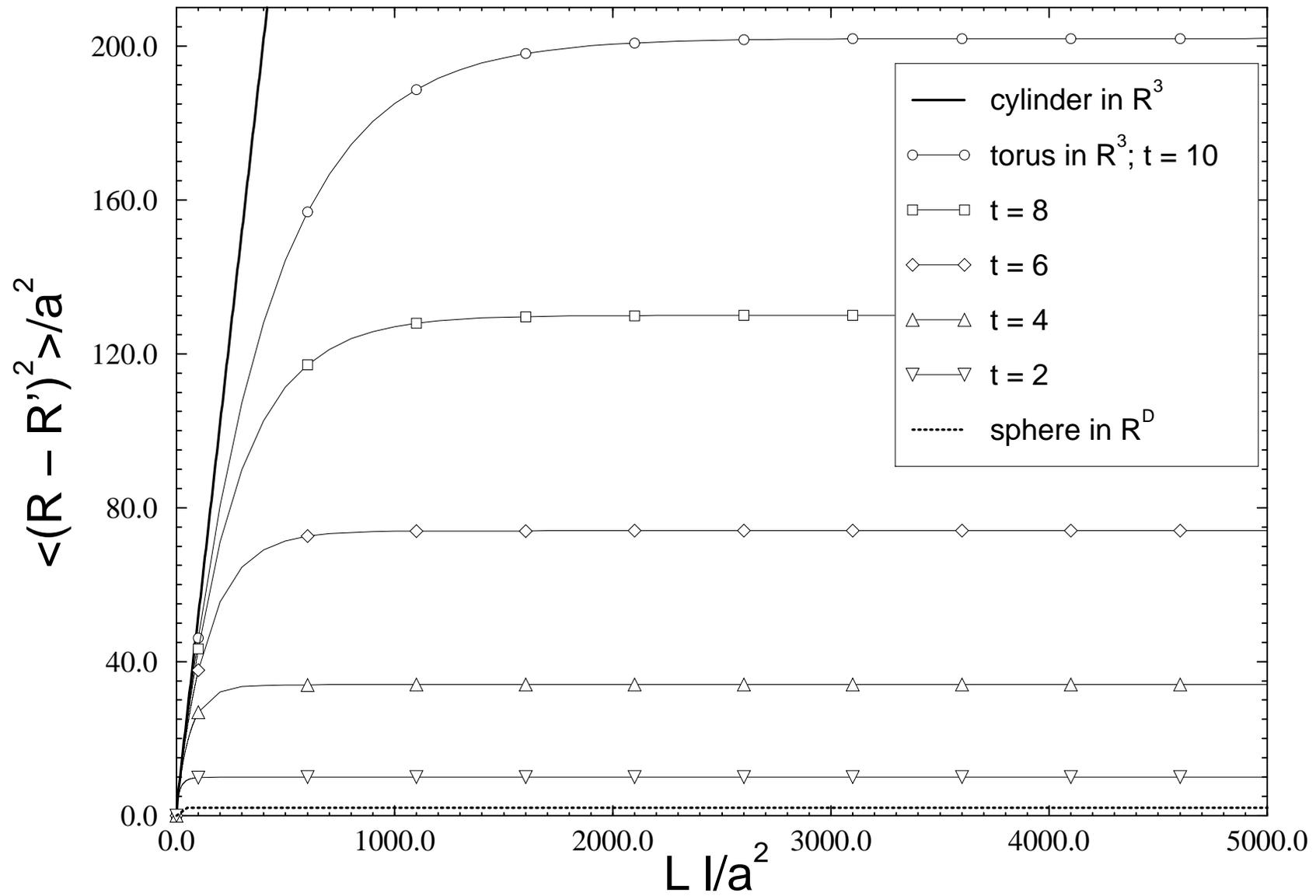
\end{document}